# Influence of the Halide Ion on the A Site Dynamics in FAPb$X_3$ ($X$ = Br and Cl)

V. K. Sharma[1,2*], R. Mukhopadhyay[1,2], A. Mohanty[3], V. García Sakai[4], M. Tyagi[5,6], D. D. Sarma[3,7]

[1]*Solid State Physics Division, Bhabha Atomic Research Centre, Mumbai 400085, India*

[2]*Homi Bhabha National Institute, Anushaktinagar, Mumbai 400094, India*

[3]*Solid State and Structural Chemistry Unit, Indian Institute of Science, Bengaluru 560012, India*

[4] *ISIS Pulsed Neutron and Muon Facility, Science and Technology Facilities Council, Rutherford Appleton Laboratory, Didcot OX11 0QX, United Kingdom*

[5]*National Institute of Standards and Technology, Center for Neutron Research, Gaithersburg, MD 20899, United States*

[6]*Department of Materials Science and Engineering, University of Maryland, College Park, MD 20742, United States*

[7]*CSIR-National Institute for Interdisciplinary Science and Technology (CSIR-NIIST), Industrial Estate P.O., Pappanamcode, Thiruvananthapuram 695019, India*

*Corresponding Author: Email: sharmavk@barc.gov.in; Phone +91-22-25594604

**Abstract**

The optoelectronic properties and ultimately photovoltaic performance of hybrid lead halide perovskites, is inherently related to the dynamics of the organic cations. Here we report on the dynamics of the formamidinium (FA) cation in FAPb$X_3$ perovskites for chloride and bromide varieties, as studied by neutron spectroscopy. Elastic fixed window scan measurements showed the onset of reorientational motion of FA cations in FAPbCl$_3$ to occur at a considerably higher temperature compared to that in FAPbBr$_3$. In addition, we observed two distinct dynamical transitions only in the chloride system, suggesting a significant variation in the reorientational motions of the FA cation with temperature. Quasielastic neutron scattering data analysis of FAPbCl$_3$ showed that in the low temperature orthorhombic phase, FA cations undergo 2-fold jump reorientations about the C-H axis which evolve into an isotropic rotation in the intermediate tetragonal and high temperature cubic phases. Comparing the results with those from FAPbBr$_3$, reveal that the time scale, barrier to reorientation and the geometry of reorientational motion of the FA cation are significantly different for the two halides. We note that this dependence of the dynamic properties of the A-site cation on the halide, is unique to the FA series; the geometry of methylammonium (MA) cation dynamics in MAPb$X_3$ is known to be insensitive to different halides.

# I. Introduction

Hybrid organic-inorganic perovskites (HOIPs) have attracted vast attention in the last one decade due to their excellent photovoltaic properties e.g., high optical absorption coefficients [1-2], high carrier mobilities, long carrier diffusion lengths [3], low exciton binding energies [4] and very high power conversion efficiencies (PCE) at certain compositions [5]. Due to their tuneable band gap and defect tolerance [6-7], they contribute towards improved PCE. These properties make them attractive for various applications besides photovoltaics [8-11], such as photodetectors [12-13], light emitting diodes (LEDs) [14-16], and lasers [17]. Within a decade, the PCE of the optimised composition has sharply increased from 3.8 % to in excess of 25 % [5, 18]. In addition, HOIPs also have many other attractive features, such as its low-cost solution-based processability which make them easy to handle. The HOIP system is generally represented as $ABX_3$, where $A$ is an organic cation, typically methylammonium (MA: $[CH_3NH_3]^+$) and/or formamidinium (FA: $[HC(NH_2)_2]^+$); $B$ is a metal ion such as $Sn^{2+}$ or $Pb^{2+}$; and $X$ is a halogen ion, namely $I^-$, $Br^-$, or $Cl^-$. Among metal ions, $Pb^{2+}$ has been used invariably in photovoltaic devices due to its higher stability and PCE compared to those with $Sn^{2+}$. These materials are structurally and compositionally distinct from traditional semiconductors (e.g., Si, CdTe), and exhibit additional dynamical degrees of freedom due to the stochastic reorientation of the organic cations, which play an important role in their photovoltaic and optoelectronic properties [19-20]. The long electronic excited-state lifetimes have been attributed to the reorientational dynamics of the organic cation through giant polaron formation [21], phonon bottleneck [22-23], or Rashba splitting–based mechanisms [24]. The reorientational dynamics of organic cations are found to influence the steady-state electronic excitation and transport properties, which have been explained based on the coupling between the orientation of organic cation and the inorganic framework [19, 25-26]. Organic cations linked with the inorganic framework through hydrogen bonding and quadrupolar interactions results in local distortions [19, 27]. A large anharmonicity has been reported and attributed to the coupling between the dynamics of the inorganic lattice and the reorientational motion of the cations [28-29]. Moreover, recent studies have shown that the rotational motion of the cations is associated with the charge carrier lifetimes [30-31]. These reports point to the importance of the reorientational dynamics of cations in determining the properties of HOIP materials, underlining the need to investigate this rather unusual and somewhat unique aspect of these hybrid materials.

The dynamics of the organic cations can be probed using various experimental methods such as nuclear magnetic resonance (NMR) [18, 32], Raman spectroscopy [29], ultrafast 2D vibrational spectroscopy [33], quasielastic neutron scattering (QENS) [19, 30, 34-38] as well as computational methods, e.g., molecular dynamics (MD) simulation [39-44]. While cation dynamics in MAPb$X_3$ have been widely investigated [34-37, 45-48], there are scarce reports of formamidium-based perovskites, FAPb$X_3$ [19, 35]. The dynamics of the FA cation in FAPb$X_3$ would be significantly differ from the MA in MAPb$X_3$ due to its various physical differences, viz. larger molecular size, distinct molecular symmetry, different dipole and quadrupolar moments, as well as due to presence of the amine groups. While the dipole moment of FA is weaker, its quadrupole moment is stronger than that of MA. In fact, contrasting dynamical behaviour of MA and FA cations have been observed in the $A$PbBr$_3$ system ($A$ = MA and FA) as studied by QENS [35]. Measurements established a plastic-crystal-like phase within the orthorhombic phase of MAPbBr$_3$, characterised by the 3-fold rotations of the MA cations around the C−N axis. In addition a 4-fold orientational motion of the MA cation about an axis perpendicular to the C-N axis is found to exist in the high temperature tetragonal and cubic phases [35]. In contrast, no such plastic-crystal-like feature was observed in FAPbBr$_3$; the FA cation was found to undergo isotropic reorientations over the entire temperature range, irrespective of its different crystallographic phases [35].

FAPb$X_3$ ($X$ = Br, Cl) have attracted considerable attention due to their better thermal stability at high temperatures [49], superior transport properties [50], and smaller bandgaps for a more efficient harvesting of the solar spectrum [51]. In our previous report [35], we have extensively aimed at understanding the FA cation dynamics for the bromide system. Here, our objective is to delineate the nature of FA cation dynamics for the chloride system and, to correlate the dynamical features of the FA cations for the different halides. Recently, chloride substitution at the halide site, has led to various studies e.g. effect of light excitation on anion segregations [52-53]. Also, incorporation of Cl$^-$ in the perovskite precursors is found to enhance the crystal size and storage stability of the solar cells [54]. Though FAPbCl$_3$, has a relatively high band gap, it is used for several applications including as a stabilizing agent in organic solar cells [55]. FAPbCl$_3$ is also used as an effective charge extracting interlayer [55]. In addition, the better stability of FA-based solar cells against humidity, is attributed to the Cl-substitution at the $X$-site [56]. Recently, FAPbCl$_3$, has been effectively used as sensitive and selective ammonia gas sensor [57]. In another study, Raman measurements have been reported for FAPbBr$_3$ and FAPbCl$_3$ to elucidate the effect of halide ion on the libration of FA and vibrations of PbX$_6$ octahedra [58]. An increase in the frequencies of the vibrational bands were observed as the more electronegative Cl⁻ ion is incorporated into the perovskite scaffold, due to stronger Cl - NH$_2$ electrostatic interactions. The aim of the present study is to directly probe the reorientational motion of the FA cation in FAPb$X_3$ ($X$ = Cl and Br) to understand the role played by the halide ions on the rotational dynamics of the organic moiety. We achieve this by investigating the reorientational motion of FA cations in FAPbCl$_3$ and compare with that in FAPbBr$_3$ across different phases, employing QENS. Neutron scattering from FAPb$X_3$ is dominated by the scattering from the cation, specifically from the hydrogen due to its large scattering cross section ($\sigma^H$ = 82 barn) compared to all other elements (e.g. $\sigma^C$ = 5.6 barn; $\sigma^N$ = 11.5 barn; $\sigma^{Pb}$ = 11 barn; $\sigma^{Br}$ = 5.9 barn; $\sigma^{Cl}$ = 16.8 barn). Being a scattering method, QENS offers an additional advantage over other complimentary techniques like NMR, and dielectric spectroscopy, by directly accessing the spatial information of the cation dynamics through an additional control variable, the wave vector transfer ($Q$). Hence, a complete set of parameters including the characteristic time scale, geometry, and activation energy associated with any molecular motion can be obtained from these measurements [59]. QENS has been previously used to study cation dynamics in MAPb$X_3$ for different halides $X$ = Cl, Br, I [34-37, 45-47]. It was found that different halides do not affect the geometry of the cation reorientation, but the activation energy and the time scale of motion are modulated [34-37, 45-47]. Our measurements suggest that not only the time scale and activation energy, but also the geometry of reorientation motion of FA, are fundamentally different between FAPbCl$_3$ and FAPbBr$_3$.

## II Experimental Methods

### A. Sample preparations

FAPbCl$_3$ was prepared inside the glove box by mixing FACl and PbCl$_2$ in an equimolar proportion in N,N-dimethylformamide, in order to make a 0.5 M solution. The solution was stirred for 30 min at RT. The precursor solution was drop cast on a glass substrate which was preheated at 120 °C, and kept for 20 minutes. The final product in the form of white-colourless powder and small crystals were scratched from the glass substrate and stored in an inert atmosphere. FAPbBr$_3$ was prepared by dissolving an equimolar quantity of FABr and PbBr$_2$ in N,N-dimethylformamide and stirred for 1 h, to get a 1 M solution. This solution was drop-cast on preheated-glass slides (at 170°C) inside a glovebox. After 20 min of heating, the orange powder and small crystals were obtained, the latter powdered, and stored in the glove box.

## B. Neutron Scattering Measurements

Elastic intensity scans (or elastic fixed window scans, EFWS) and QENS measurements were carried out to investigate reorientational motions of the FA cations in FAPb$X_3$ ($X$ = Br and Cl) samples.

### i) EFWS Measurements

EWFS measurements were carried out on FAPbCl$_3$ and FAPbBr$_3$ using the High Flux Backscattering Spectrometer (HFBS)[60] at the NIST Center for Neutron Research (NCNR), USA. It should be noted that high energy resolution back-scattering spectrometers at reactor sources like the one used here are an optimal choice for collecting the energy-resolved elastic scattering. HFBS uses a Si(111) monochromator and analyzer crystals, which owing to its characteristic $d$-spacing of 6.27 Å achieves an energy resolution of 0.8 μeV (Full Width at Half Maximum (FWHM)). The useable $Q$ range was 0.36 – 1.75 Å$^{-1}$. EFWS measurements were carried out in the temperature range of 4-295 K with a ramp rate of 0.8 K/min. At each temperature, counting time was set to 1 min, thus providing a temperature resolution of 0.8 K. EFWS measurements were carried out in both the heating and cooling cycles to investigate the reversibility and hysteresis involved of the observed phenomena. For standard data reduction, DAVE software [61] version 2.5 was used.

### ii) QENS measurements

QENS measurements were carried out on FAPb$X_3$ ($X$ = Br and Cl) using the IRIS near backscattering spectrometer at the ISIS facility, UK [62]. IRIS is a time-of-flight backscattering inverted geometry spectrometer which uses PG002 crystals to analyse a fixed final neutron energy of $E_f$ = 1.84 meV. In this configuration, the spectrometer has an energy resolution of 17.5 μeV (FWHM) and the available energy transfer range, $E = E_i - E_f$ is ± 0.5 meV. The accessible $Q$ range in this configuration was 0.6 – 1.8 Å$^{-1}$. Detectors encompassing coherent Bragg reflections in both samples were removed while reduction of data. Previous results on the Br variety [35] were reported based on the measurements on HFBS ($\Delta E$ = 0.8 μeV) and the FOCUS spectrometer ($\Delta E$ = 45 μeV), at SINQ, PSI, Switzerland. Here we add measurements on IRIS ($\Delta E$ = 17.5 μeV) to facilitate a direct comparison of the two systems, $X$ = Br and Cl, on spectrometers of the same energy resolution. QENS spectra were recorded for both systems at different temperatures in the range 150 - 320 K. By virtue of a small number of diffraction detectors available on IRIS (at a scattering angle ~ 170º), diffraction patterns were also recorded to monitor the variations in structure simultaneously with the QENS data. The samples were filled in an aluminum sachet, which was wrapped up in an annular aluminum can. A sample thickness of ~ 0.2 mm was chosen such that the scattering from the samples is no more than 10 %, thereby minimizing multiple scattering effects. The instrument resolution was obtained by measuring the spectra from a standard vanadium sample. For data reduction, MANTID software [63] version 3.13 was used.

**III Data Analysis**

In a QENS experiment from a hydrogenous system, the observed scattered intensity can be expressed in terms of the incoherent dynamic structure factor, $S(Q,\omega)$ which in general can be written as [59]

$$S(Q,\omega) = A(Q)\delta(\omega) + [1 - A(Q)]L(\Gamma,\omega) \quad (1)$$

where $\omega$ is the angular frequency corresponding to the energy transfer. The first term in Eq. (1) is the elastic term and the second term is the quasielastic one. The contribution of the elastic scattering to the total scattering is known as the elastic incoherent structure factor (EISF) which provides information about the spatial extent and the geometry of the molecular motions. Hence, $A(Q)$ in Eq. (1) corresponds to the EISF. $L(\Gamma,\omega)$ is a Lorentzian function with $\Gamma$ as the half width at half maximum (HWHM) which is inversely proportional to the time scale of the motion. Eq. (1) is convoluted with the instrumental resolution function, and least squares fitting was used to describe the observed QENS spectra for FAPb$X_3$ ($X$=Cl & Br) and the two parameters, EISF ($A(Q)$) and $\Gamma(Q)$ are determined. The analysis of these parameters provides information towards the geometry and time scale of the molecular motion of FA cation.

As the molecular configuration of the FA molecule is very different from the MA, the dynamical modes in the FA compound are expected to be very different. In FA motions around the C-N axis are not expected like in MA, since FA is planar, due to the partial double-bond character of the C-N bonds [64]. The plausible reorientational motions of the FA cation could include, jump rotation about the C-H axis, uniaxial rotational diffusion (URD) around the N-N axis and isotropic rotational diffusion (IRD). It may be noted that IRD is the effective model and it happens when both the rotations around the C-H axis and N-N axis are rapid in the scale of the measurements, making the molecule appear as if it were rotating isotropically. Schematics of these models are shown in the Fig. 1. The radii of rotation for each model can be calculated from the molecular configuration of the FA molecule and are 1.89, 0.99 and 1.8 Å for the jump rotation about the C-H axis, URD around the N-N axis and IRD, respectively.

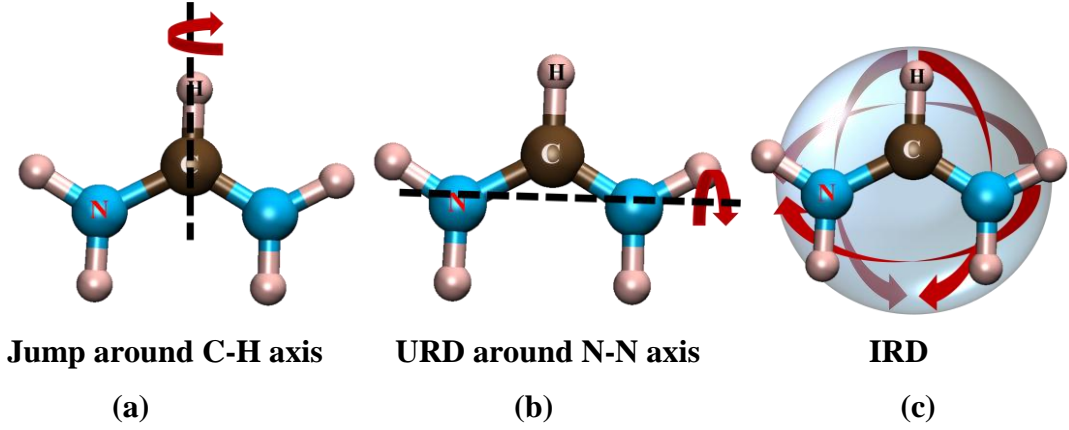

| Jump around C-H axis | URD around N-N axis | IRD |
| :---: | :---: | :---: |
| (a) | (b) | (c) |

Fig. 1 Schematic of various possible rotational motions of FA cations, (a) jump rotation about the C-H axis, (b) uniaxial rotational diffusion (URD) about N-N axis and (c) isotropic rotational diffusion (IRD).

As shown previously [34-35, 45-46], it is possible that for a given temperature all the cations are not mobile. Hence, general expressions for the EISF for fractional 2-fold, 4-fold jump rotation about the C-H axis, URD around N-N axis and IRD can be written as [59, 65]

$$[A(Q)]_{2-fold} = p_x + \frac{1}{2}(1-p_x)\left[1 + j_0(2QR_{CH})\right] \quad (2)$$

$$[A(Q)]_{4-fold} = p_x + \frac{1}{4}(1-p_x)\left[1 + 2j_0(QR_{CH}\sqrt{2}) + j_0(2QR_{CH})\right] \quad (3)$$

$$[A(Q)]_{URD} = p_x + \frac{1}{N}(1-p_x)\sum_{i=1}^{N} j_0\left(2QR_{NN}\sin\left(\frac{i\pi}{N}\right)\right) \quad (4)$$

$$[A(Q)]_{IRD} = p_x + (1-p_x)j_0^2(QR_{iso}) \quad (5)$$

where $p_x$ is the fraction of the immobile cations and $j_0$, is the spherical Bessel function of zeroth order. $R_{CH}$, $R_{NN}$, $R_{iso}$ are the radii of rotation for the jump rotation about the C-H axis, URD around the N-N axis and IRD, respectively. It may be noted the values of the fraction of mobile cations is an average value associated with a specific model.

To obtain the characteristic times quantitatively, it is important to analyze the HWHM, $\Gamma$, using the appropriate scattering laws. The corresponding scattering laws for these models can be found in Refs. [59] and [65]. Among the different models mentioned above, two different motions, namely, the fractional 2-fold jump rotation and IRD are found to best describe the QENS data over different temperature ranges. The dynamic structure factor for the fractional 2-fold jump rotation of FA around the C-H axis, can be written as[59]

$$S(Q,\omega) = p_x\delta(\omega) + (1-p_x)\frac{1}{2}\left[\{1+j_o(2QR_{CH})\}\delta(\omega) + \frac{1}{\pi}\{1+j_o(2QR_{CH})\}\left\{\frac{2/\tau_{2f}}{(2/\tau_{2f})^2+\omega^2}\right\}\right] \quad (6)$$

where $\tau_{2f}$ is the characteristic relaxation time for the 2-fold rotational motion. From the above equation, it is clear that HWHM of the Lorentzian for the fractional 2-fold jump rotation would be $2/\tau_{2f}$.

The dynamic structure factor for the fractional IRD model can be written as[65]

$$S(Q,\omega) = p_x\delta(\omega) + (1-p_x)\left[j_0^2(QR_{iso})\delta(\omega) + \frac{1}{\pi}\sum_{l=1}^{\infty}(2l+1)j_l^2(QR_{iso})\frac{l(l+1)\tau_{iso}}{(l(l+1))^2+\tau_{iso}^2\omega^2}\right] \quad (7)$$

where $j_l$ is the spherical Bessel functions of the $l^{th}$ order and $\tau_{iso}$ is the characteristic time scale of isotropic rotation. Unlike for the 2-fold rotational motion, there is no analytical expression for the HWHM of the IRD model. However, it can be numerically calculated using Eq. (7) and $\tau_{iso}$ is obtained using least squares fitting with the measured $S(Q,\omega)$.

## IV Results and Discussion

Govinda et al.[66] have reported that $FAPbCl_3$ undergoes two first order phase transitions at about 258 K and 271 K as observed in the DSC measurements. These transitions were found to be reversible with a small hysteresis. The same study[66] also showed that at low temperatures (200 K and 100 K) the system exists in the orthorhombic (O) structure between 100 and 200 K and in the cubic (C) structure at room temperature (295 K). The neutron diffraction patterns for $FAPbCl_3$ recorded in the present experiment at different temperatures (crystal phases) are shown in Fig. 2. Three distinct phases could be identified with phase transitions appearing in the temperature ranges of 250 - 260 K for O to an intermediate phase and 270 - 280 K for the intermediate phase to C phase; consistent with the x-ray diffraction and DSC results[66]. Unfortunately, the intermediate phase is not well-characterized in the literature in terms of its phase identification. While the limited range does not allow a full refinement of the diffraction data obtained from the IRIS spectrometer, it was possible to index the pattern correspond to the intermediate phase assuming it to be a tetragonal structure, as shown in Fig. S1 (supporting information). In view of this and for the sake of convenience, we refer to this intermediate phase as tetragonal in this manuscript.

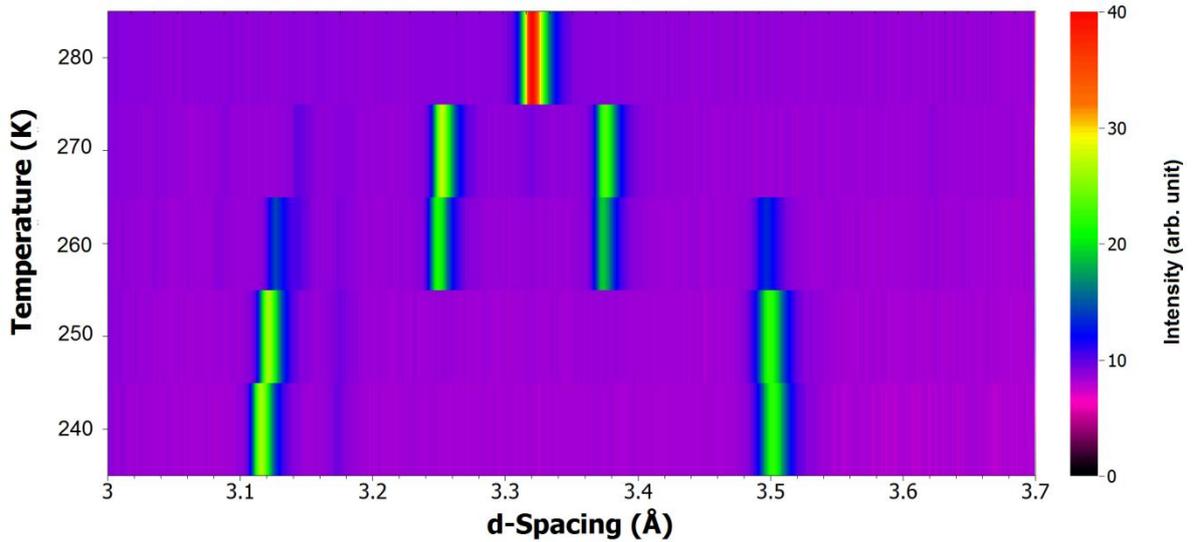

Fig. 2 The diffraction pattern for FAPbCl$_3$ recorded using IRIS spectrometer, confirming three distinct crystallographic phases; (1) orthorhombic at 250 K and below, (2) tetragonal at 260 – 270 K and (3) cubic at 280 K and above. The diffraction data were recorded simultaneously with the QENS measurements.

We carried out EFWS measurements on FAPbCl$_3$ upon heating as well as cooling. The $Q$-averaged elastic scan data in both heating and cooling cycles are shown in Fig. 3(a). For direct comparison, the $Q$-averaged elastic intensity scan for the FAPbBr$_3$ as obtained with the same experimental set up used for the chloride sample is shown in Fig. 3(b) ; the data in Fig. 3(b) is similar to the one reported earlier [35]. Unlike FAPbCl$_3$, structures of FAPbBr$_3$ at different temperatures have been studied in detail and show that phase transitions are more complex and contentious [66-68]. As discussed in Ref. [35], the different phases observed in FAPbBr$_3$ can be summarised as, (1) below ∼110 K, it has either an orthorhombic [67-68] or a trigonal [66] structure, (2) at temperatures higher than ∼180 K, it exists in the cubic[66-68] and pseudo-cubic-tetragonal [67-68] structures, and (3) a complex behavior with multiple phase transitions appears in the temperature interval between 110 and 180 K. The characteristic transition temperatures for both compounds are indicated by dashed vertical lines which allow association of changes in the elastic intensity with temperature and crystallographic phase. The elastic intensity for FAPbCl$_3$ slowly decreases initially on heating until 160 K, exhibiting typical behaviour due to thermally assisted Debye Waller factor. Then there is a rapid decrease in the elastic intensity in the temperature range 160 – 225 K corresponding to the onset and evolution of the reorientation of the FA cations. In the case of

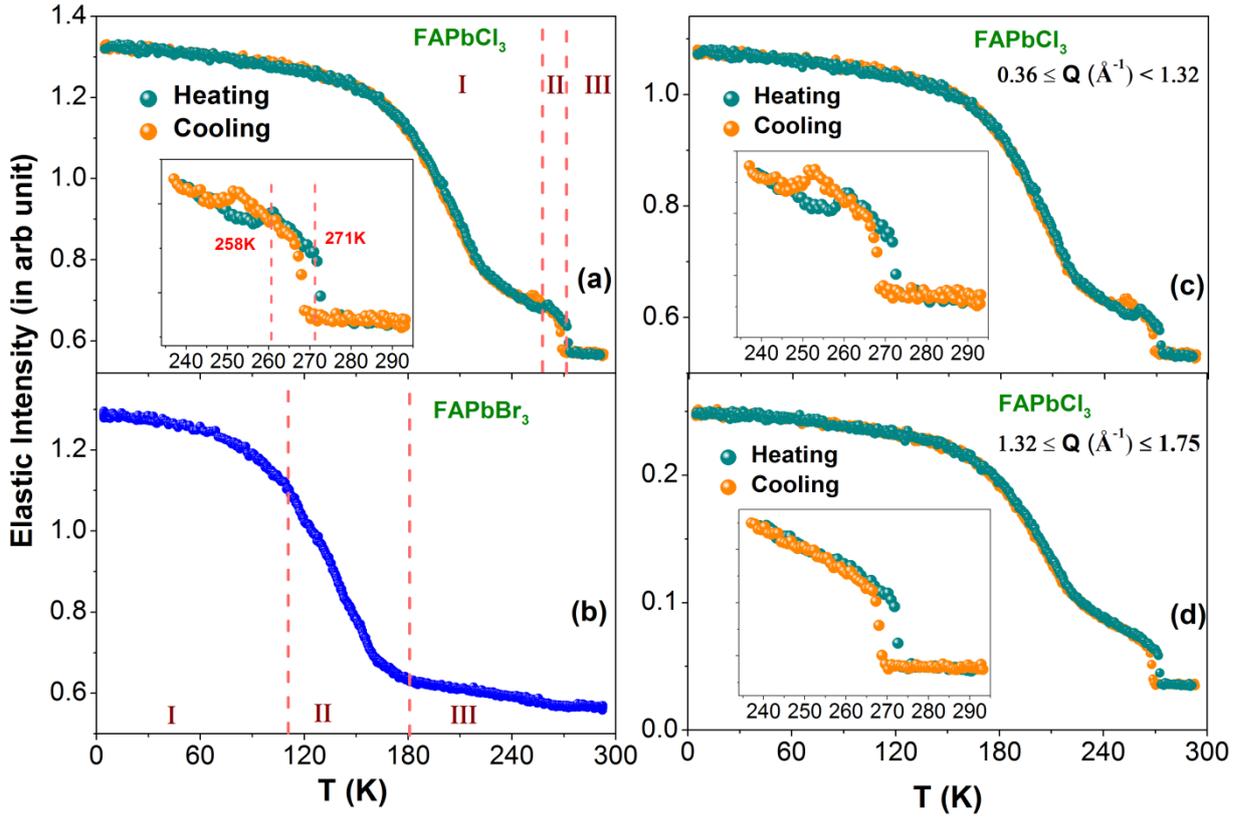

Fig. 3 (a) $Q$-averaged ($0.36 \leq Q$ (Å$^{-1}$) $\leq 1.75$) elastic intensity scans or EFWS for FAPbCl$_3$ in the heating and cooling cycles. Crystallographic phase transitions are marked by vertical dashed lines. Inset shows the elastic intensity in the expanded scale around the phase transition temperatures (235 - 295 K). The abrupt drop in the elastic intensity at 271 K upon heating corresponds to the T to the C phase transition, and the feature observed at 258 K is associated with the O to T phase transition. (b) $Q$-averaged ($0.36 \leq Q$ (Å$^{-1}$) $\leq 1.75$) elastic intensity scan data from FAPbBr$_3$ shown for direct comparison. The featureless behaviour is evident in comparison to the chloride sample. Elastic intensity scans for FAPbCl$_3$ in two different $Q$ regimes averaged over (c) $0.36 \leq Q$ (Å$^{-1}$) $< 1.32$ and (d) $1.32 \leq Q$ (Å$^{-1}$) $\leq 1.75$ chosen from the whole $Q$-set, showing that the peak like feature at 258 K is prominent only in the low $Q$ range.

FAPbBr$_3$, the onset occurs at a considerably lower temperature of about 70 K, which suggests that the activation energy for reorientational motions of the FA cation is much lower than for FAPbCl$_3$. In FAPbCl$_3$, the elastic intensity exhibits an unusual peak-like feature at ~258 K, with a small thermal hysteresis which suggests that it is related to a first order phase transition; this is consistent with earlier DSC measurements[66]. This is further investigated by separating the elastic intensity data in two different $Q$ regimes (i) from 0.36 to 1.32 Å$^{-1}$ and (ii) from 1.32 Å$^{-1}$ to 1.75 Å$^{-1}$ as shown in Figs. 3(c) and 3(d). The peak only features at low $Q$ and it originates from a manifestation of the change in the geometry of the reorientational motion of the FA cations, as will be described later from the detailed analysis of the QENS data. The sharp decrease in the elastic intensity at 270 K coincides with the 2$^{nd}$ transition observed in DSC [66], suggesting that both phase transitions are associated with changes in the microscopic dynamics of the A-site organic cation. In contrast, no apparent dynamical transition, signalled by a sharp change, was observed in the elastic intensity scan of FAPbBr$_3$.

We investigated the differences in the organic cation dynamics in these two systems with the help of QENS experiments using IRIS spectrometer (ΔE = 17.5 μeV) keeping all experimental parameters identical. Typical QENS spectra from FAPbCl$_3$ at $Q$ = 1.2 Å$^{-1}$ at different temperatures are shown in Fig. 4(a). The instrumental resolution as obtained from the measurement of the spectra from a standard vanadium sample is also shown in the figure by a dashed line. To visualise variations in the quasielastic broadenings at different temperatures, the spectra are normalised at the peak, $S(Q, 0)$. The presence of significant quasielastic broadenings at 240 K and above suggests presence of stochastic motion of the FA cation. No significant quasielastic broadening could be observed below 240 K on IRIS (ΔE = 17.5 μeV); for example, data obtained at 220 K coincide with the resolution of the instrument as shown in the Figure. It may be noted that the elastic intensity started to decrease significantly at a much lower temperature (~160 K), while significant quasielastic broadening could be observed in the QENS data obtained with IRIS spectrometer only at and above 240 K. So, the possibility of existence of low energy excitations (viz. NH$_2$ rotations) cannot be ruled out. A large jump in quasielastic broadening is observed ongoing above 270 K. This suggests a major change in the dynamics above 270 K, which is consistent with the observed sudden drop in the elastic intensity at ~271 K. For a direct comparison of the observed data, QENS spectra obtained from FAPbBr$_3$ are shown in Fig. 4(b). It is evident that the temperature evolution of the quasielastic broadening in FAPbBr$_3$ is quite different than for FAPbCl$_3$. Three major contrasting features that could be identified are, (i) absence of any sudden jump in quasielastic broadening with temperature in FAPbBr$_3$ (ii) the increase in quasielastic broadening is almost monotonous with temperature in FAPbBr$_3$ and, (iii) a measureable/significant quasielastic broadening is observed at relatively lower temperatures in FAPbBr$_3$ compared to FAPbCl$_3$. To understand this contrasting rotational dynamics of the same organic cation in the two different halide samples, QENS data analysis has been carried out in detail. Scattering law as given in Eq. (1), which has a elastic and a quasielastic component was used to describe the observed data. It was found that the data could be described very well for all the $Q$ values and at all temperatures. Typical fitted QENS spectra for FAPbCl$_3$ at $Q$ = 1.2 Å$^{-1}$ at three different temperatures, namely, 250 K, 270 K and 300 K are shown in Fig. S2. Presence of significant fraction of elastic component and $Q$ independent HWHM of the Lorentzian function suggests that the observed dynamics is localized in nature.

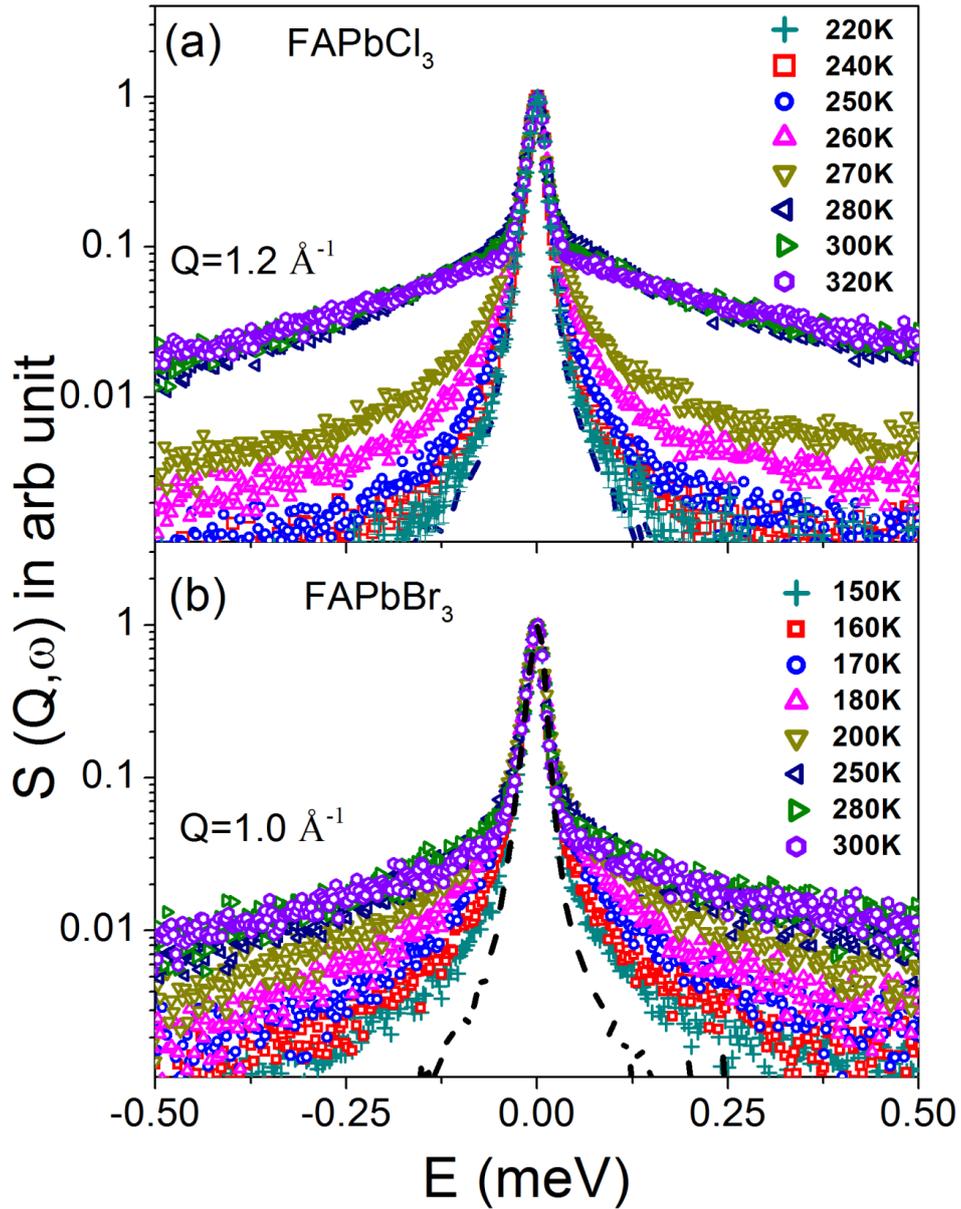

Fig. 4 Typical observed QENS spectra for (a) FAPbCl$_3$ and (b) FAPbBr$_3$ as recorded using IRIS spectrometer at different temperatures across the different crystallographic phases. QENS spectra at each temperature are normalised to peak amplitudes which enables a direct comparison of the quasielastic broadening at different temperatures. Instrument resolution as measured using standard vanadium sample is shown by dashed line in both panels.

The behaviour of the EISF with $Q$ obtained for FAPbCl$_3$ and FAPbBr$_3$ at different temperatures is shown in Figs. 5(a) and 5(b), respectively. The EISF shows very different temperature dependencies in the two systems. For FAPbBr$_3$, the EISF shows a qualitatively similar $Q$-dependence at all temperatures with a gradual progression with temperature. However, this is not the case for FAPbCl$_3$. The temperature dependent EISF of FAPbCl$_3$ shows interesting non-monotonic behaviour that can be divided into three separate regimes, ≤ 250 K, 260 - 270 K and ≥ 280 K, demarcating the different crystallographic phases, namely O, T and C phases, respectively, with distinctive features and with little temperature-dependence within each of the phases. The lowest temperature O phase shows a qualitatively different $Q$ dependence of the EISF with a minimum at 1.2 Å$^{-1}$. With an increase in the temperature across the O to T transition, the EISF shows unusual changes. While the EISF is expected to decrease with increasing temperature due to enhanced dynamics at higher temperatures, the EISF in the T phase is surprisingly higher in the $Q$ ≤ 1.32 Å$^{-1}$ regime compared to that in the lower temperature O phase (see Fig. 5(a)). This suggests some qualitative changes in the reorientational motion of the FA cation across the O to T transition. The EISF of FAPbCl$_3$ shows another abrupt change across the phase transition from the intermediate T phase to the high temperature C phase. However, the nature of the $Q$-dependence of the EISF is found to be similar in both intermediate T and high temperature C phases, decreasing monotonically with $Q$. The qualitative changes in the $Q$-dependencies of the EISF of FAPbCl$_3$ in the different temperature regimes clearly indicate that the nature of the rotational dynamics of FA cations in the low temperature O phase is significantly different than that in the intermediate T and high temperature C phases of FAPbCl$_3$. It is also evident that the EISF in the intermediate T phase shows an unusual behaviour in terms of an increase, instead of the usual decrease, with increasing temperature while moving from the lower temperature O phase to the intermediate T phase. This behaviour of EISF can be understood based on the EFWS results which showed that an unusual peak like feature in the elastic intensity scan at O phase to T phase is mainly associated with the low $Q$ aspects ($Q$ < 1.32 Å$^{-1}$). If one observes the behaviour of the EISF in the O phase and T phase, it is very clear that the EISF is higher in the intermediate T phase for $Q$ < 1.32 Å$^{-1}$. This explains the

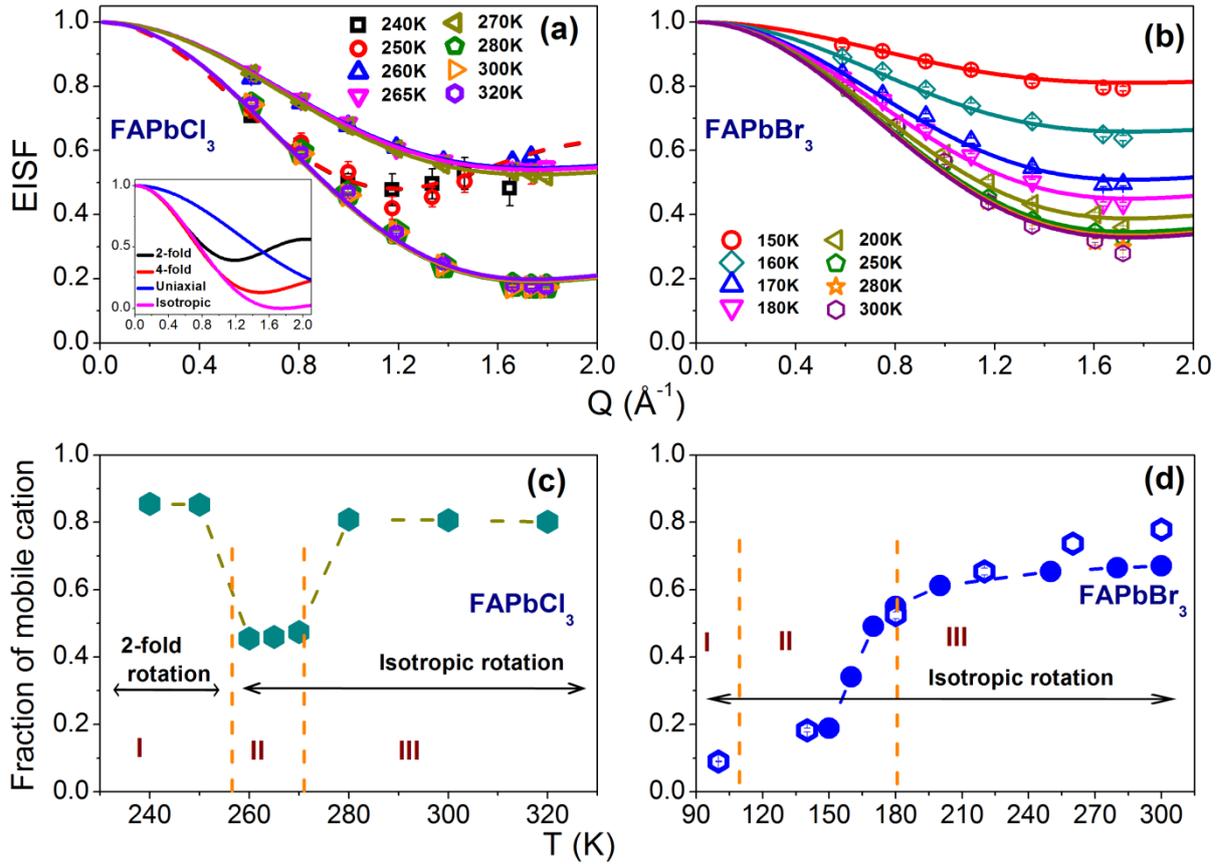

Fig. 5 (a) EISF for FAPbCl$_3$ at different temperatures across three different crystallographic phases. Solid and dashed lines are the fits assuming fractional 2-fold jump rotation about the C-H axis (240-250K, in the O phase) and fractional isotropic rotational diffusion (260-270 K, in the T phase and 280-320 K, in the C phases) models, respectively. The inset shows the theoretical EISF calculated for different plausible rotational motions of the FA cation, namely 2-fold and 4-fold jump rotations about C-H axis, uniaxial rotation about N-N axis and isotropic rotation assuming all cations are mobile. (b) EISF for FAPbBr$_3$ at different temperatures. Solid lines are the fits assuming a fractional isotropic rotational diffusion model. (c) Variation of the fraction of mobile FA cations in FAPbCl$_3$ and (d) in FAPbBr$_3$, with temperature. Fractions of mobile FA cations in FAPbBr$_3$ obtained earlier [35] are also shown by open circles.

observed peak-like feature in the elastic scan data of FAPbCl$_3$. The sudden change in EISF or in other words the abrupt alteration in the geometry of reorientational motion of the FA cation manifests to the observed peak-like feature observed in the elastic scan. It may be noted this is very specific to the system, correlated to the different phases and associated dynamical mode of the species involved.

Various plausible models for the reorientation motion of the FA cation are 2-fold and 4-fold jump rotations about the C-H axis, URD around N-N axis and IRD as shown in Fig. 1. The theoretically calculated EISF for these models are shown in the inset of Fig. 5. It is evident that the EISF calculated for the models in which the FA cation undergoes only 2-fold or 4-fold jump rotational motion around the C-H axis, is very similar to that observed in the high temperature regime of the O phase of $FAPbCl_3$: first a decrease, reaching a minimum at $Q \sim 1.2$ Å$^{-1}$, before increasing again with increasing $Q$. The behaviour of the observed EISF in the T and *C* phases suggests either IRD or URD motion. Least squares fittings have been carried out using all these models (Eqs. 2-4) with the fraction of immobile cations ($p_x$) as a parameter, keeping the radius of rotations fixed to the calculated values (mentioned earlier in section III) corresponding to that specific model. It is found that in the O phase, 2-fold jump rotations around the C-H axis describe the data best, as shown in Fig. S3 for 250 K. However, IRD describes the data best in the T and C phases as shown for 265 K (T phase) and 300 K (C phase) in Figs. S4 and S5, respectively. Solid and dashed lines in Fig. 5(a) correspond to the fits as per the fractional IRD and 2-fold rotation around the C-H axis, respectively. Therefore, it is clear that the FA cation in $FAPbCl_3$ changes its mode of reorienational motion at the transition temperature (~260K), from a 2-fold reorientation to IRD when the system undergoes a first order transition from the O phase to the T phase. The transition from the T phase to the C phase the nature of reorientational motion does not undergo a qualitative change, still being consistent with the IRD model; however, in quantitative terms the reorientational motions become much more rigorous, as indicated by a more rapid decrease of EISF (Fig. 5(a)). This reflected as increased fraction of mobile cations for the higher temperature cubic phase as shown in Fig. 5(c). In the O phase, the fraction of mobile cations is found to be 0.88, which means that about 88 % of the FA cations are mobile and performing 2-fold jump rotation around the C-H axis. In the T phase, the nature of the rotational motion changes, presumably because of the ability of this structure to support URD around the N-N axis that was suppressed in the low temperature O phase; this URD coupled with the facile rotations around the C-H axis, together give rise to what appears as an isotropic rotation describing the EISF best with a fraction of mobile cations of about 46%. In the C phase, the fraction of mobile cation is found to be much higher (81%) due to a higher fraction of FA ions being able to execute URD in conjunction with the rotations around the C-H axis.

It is interesting to note that within a crystallographic phase, the fraction of mobile cations does not change with temperature (Fig. 5(c)), consistent with the relative independence of EISF with temperature within a given phase. Our earlier work [35] on FAPbBr$_3$ established that the IRD model could describe the data well over the entire temperature range including the low temperature O phase. More extensive data from FAPbBr$_3$ in the present set of experiments also could be described well with the IRD model as shown in Fig. 5(b). The obtained fractions of mobile cations are shown in Fig. 5(d). Temperature regions corresponding to different crystallographic phases, as discussed earlier, are also marked in the Figs. 5(c) and 5(d) by dashed vertical lines and Roman numerals. It is clear that an increase in the fraction of mobile cations in region II is mainly responsible for the rapid decrease in the elastic intensity in the EFWS in this temperature range. These results clearly demonstrate the contrast in the FA cation dynamics in the two halide systems. The basic difference between the rotational dynamics found in the two compounds appears to arise from the ability of the FA ions to rotate in an isotropic manner right from the lowest temperature orthorhombic phase in case of the bromide compound, whereas the isotropic rotation sets in for the chloride compound at a higher temperature only after the system makes a transition out of the orthorhombic phase. Since the IRD is possible when rotations around both C-H and N-N axis are facile, it would appear that specifically the URD around the N-N axis is hindered in the orthorhombic phase for the chloride compound and this rotation becomes step-wise accessible with increasing temperature through successive phase transitions exhibited by FAPbCl$_3$. It is tempting to correlate the ability/inability of FA rotations around the N-N axis in the two compounds to the structural data. The bromide compound with the larger Br$^-$ ion has larger lattice parameters and unit cell volume compared to those of the chloride compound. Specifically, the unit cell volumes of FAPbBr$_3$ and FAPbCl$_3$ are 827.72 Å$^3$ and 734.09 Å$^3$, respectively, in the orthorhombic phases.[66, 69] The considerable reduced unit cell volume of the chloride compound is the cause of hindering the URD of FA around its N-N axis. Similarly, unit cell volumes of FAPbBr$_3$ and FAPbCl$_3$ in the high temperature cubic phase are 215.58 Å$^3$ and 188.91 Å$^3$, respectively [66, 69]. While the higher temperature in the phase stability region of the cubic phase ensures IRD for the FA ions, we shall see later that the smaller unit cell volume leads to a higher activation energy for the case of FAPbCl$_3$.

As mentioned earlier, the variation of the obtained HWHM, $\Gamma$, is nearly independent of $Q$, which is a characteristic of a localized motion. The temperature dependence of the $Q$-averaged quasielastic widths, $\Gamma$, for FAPbCl$_3$ and FAPbBr$_3$ are plotted in Figs. 6(a) and 6(b) respectively. It is evident that for both the systems, $\Gamma$ increases with an increase in the temperature as expected for a thermally activated stochastic motion. Two major differences are evident between FAPbCl$_3$ and FAPbBr$_3$: (i) in FAPbCl$_3$, Γ increases discontinuously at ~ 270 K where the transition from T to C phase transition, whereas no discontinuity is observed in FAPbBr$_3$, and (ii) at any given temperature, a relatively lower value of Γ in FAPbCl$_3$ suggesting slower cation dynamics compared to FAPbBr$_3$. These are consistent with the elastic intensity scan data which showed a sudden drop in FAPbCl$_3$ at the T to C phase transition, while no such abrupt drop in the elastic intensity is observed in the case of FAPbBr$_3$. The time scale for the 2-fold reorientational motion of the FA cations is obtained using $\tau_{2f} = 2/\Gamma$ as per the scattering law given in Eq (6). While for IRD, Eq. (7) is used to describe the behaviour of the HWHM and the characteristic relaxation times that are obtained using least square fitting. Typical fits of HWHM for FAPbCl$_3$ in different phases are shown in Fig. S6. It is evident that the IRD model could describe the HWHM behaviour in FAPbCl$_3$ in both T and C phases. Eq. (7) has also been used to describe the HWHM for FAPbBr$_3$ over the whole temperature range. Activation energies for the different phases are estimated from the analysis of temperature dependence of the relaxation time. Arrhenius plots of the obtained relaxation time for FAPbCl$_3$ and FAPbBr$_3$ are shown in Figs 6(c) and 6(d), respectively. Solid lines are the fits as per the Arrhenius law $\tau \propto \exp(E_a/k_B T)$ where $E_a$ is the activation energy, and $k_B$ is the Boltzmann constant. For FAPbCl$_3$, activation energies obtained in the O, T and C phases are 220 ± 31 meV, 345 ± 47 meV and 82 ± 1 meV, respectively. It may be noted that in the O phase of FAPbCl$_3$, the obtained activation energy corresponds to the 2-fold rotational motion of the FA cation and in the T and C phases, they correspond to IRD model. The $E_a$ in the T phase is very large and consistent with the fraction of mobile cations being the lowest, indicating a smaller number of cations are contributing to the observed dynamics. As the system goes to the cubic phase, the activation energy for the cation rotation decreases significantly to 82 ± 1 meV and also the mobile fraction is very high. This is no surprise; the temperature is high and the system is in the most symmetric cubic phase. In case of FAPbBr$_3$, it is found that the activation energy for FA cations in the high temperature C phase (region III) and intermediate region (II) are 51 ± 5 meV and 39 ± 7 meV, respectively. It is evident that activation energies for FA cations in FAPbBr$_3$ are much smaller than those in FAPbCl$_3$. It may be noted that in the high temperature C phase, the molar volume of FAPbCl$_3$ is lower than that of FAPbBr$_3$ by more than 13% [66], explaining the higher activation energy for the Cl sample. While comparing with the low temperature O phase, $E_a$ in FAPbCl$_3$ is found to be much higher than that observed in FAPbBr$_3$ (106 meV) [35]. This is also consistent with the fact that onset of the motion in FAPbCl$_3$ occurs at a much higher temperature than in FAPbBr$_3$. It may be noted that the unit cell volume for the chloride sample is also lower than Br in the O phase [66, 69]. This is in stark contrast with the MAPbX$_3$ (X = Cl, Br and I), in which it was found to follow $(E_a)_{Cl} < (E_a)_{Br} < (E_a)_{I}$ in the O phase[34-35, 45-46]. For the MA cation in MAPbX$_3$, the activation energy in the O phase is found to be substantially lower compared to FA cation in FAPbX$_3$. For example, it is found to be 26.7 ± 0.9 meV for MAPbBr$_3$ [35] which is much lower than for FA in FAPbCl$_3$ (220 meV) and FAPbBr$_3$ (106 meV) [35].

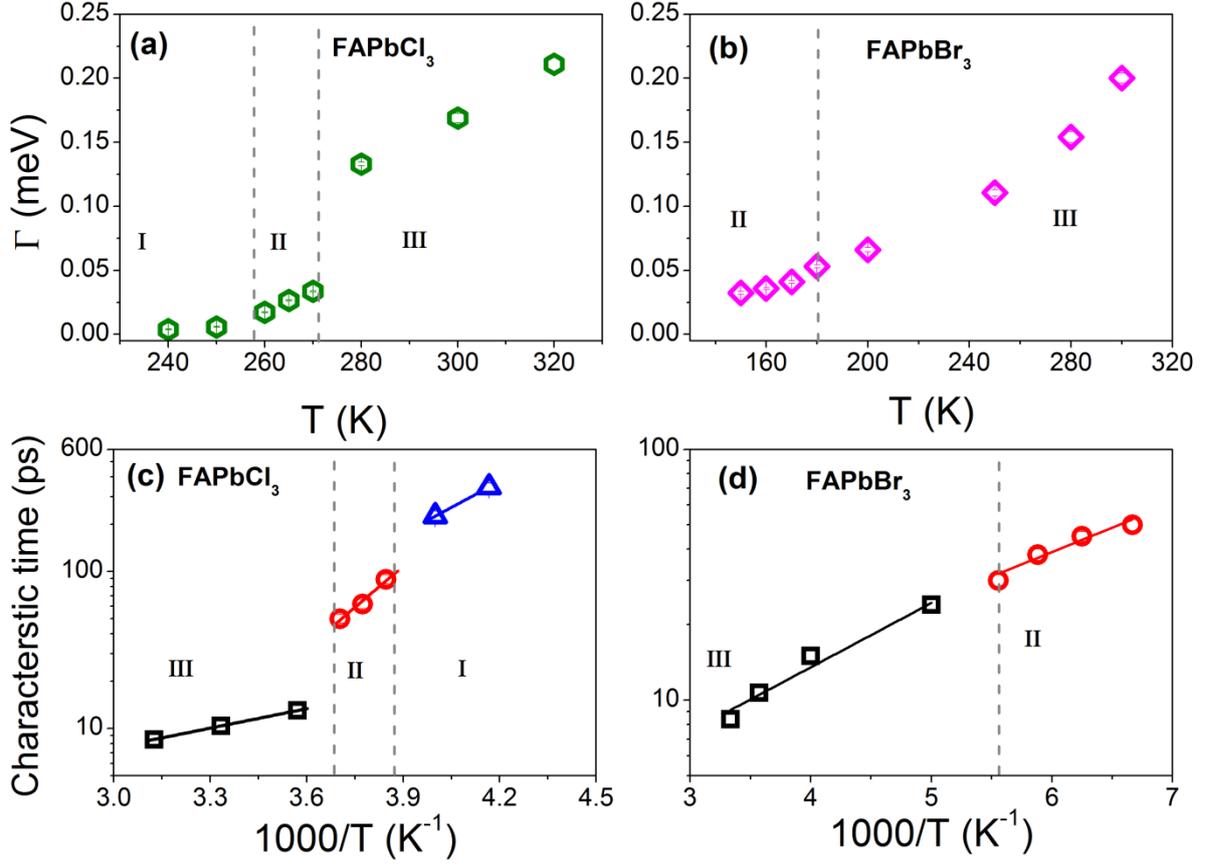

Fig. 6 Evolution of HWHM with temperature in (a) FAPbCl$_3$ and (b) FAPbBr$_3$. In case of FAPbCl$_3$, HWHM varies discontinuously at both the transition points, at 258 K and 271 K correspond to O to T and T to C phase transitions. Characteristic times for the rotational motion of the cation as obtained from the fits of the HWHM for (c) FAPbCl$_3$ and (d) FAPbBr$_3$. Different regions involving different activation energies are evident in the two systems.

# V Conclusions

Neutron scattering experiments reveal considerable differences between the reorientational dynamics of FA cations in $FAPbCl_3$ and $FAPbBr_3$. The reorientational motion of the FA cation in $FAPbCl_3$ sets in at higher temperature compared to $FAPbBr_3$, and exhibits two distinct dynamical transitions, unlike the bromide case. Detailed analysis of the neutron scattering data established that, for $FAPbCl_3$, in the high temperature regime within the low temperature orthorhombic (O) phase, FA cations undergo 2-fold jump reorientation about the C-H axis, which changes to an isotropic rotation in the higher temperature tetragonal (T) and cubic (C) phases. In contrast, the isotropic rotational diffusion has been found to describe the cation dynamics in $FAPbBr_3$ over the whole temperature range encompassing different crystallographic phases. The Arrhenius plots clearly show three different regions of dynamics in $FAPbCl_3$, associated with different activation energies. In the low temperature O phase, the activation energy ($E_a$) is found to be 220 ± 31 meV. While the dynamical mode of the cations changed from 2-fold reorientation to IRD in the intermediate T phase, $E_a$ is found to be 345 ± 47 meV, and in the high temperature C phase it reduces to 82 ± 1 meV. In contrast for $FAPbBr_3$, $E_a$ of 51 ± 5 meV is found in the high temperature C phase and 39 ± 7 meV in the intermediate region (region-II). In the low temperature O phase, $E_a$ for $FAPbBr_3$ was found to be 106 meV [35], approximately half of that in the O phase of $FAPbCl_3$. Therefore, the activation energy for the cation dynamics is much smaller in the bromide compound compared to the chloride one in each temperature regime. This enhanced activation energy in $FAPbCl_3$ is possibly due to the smaller molar volume of $FAPbCl_3$ compared to that of $FAPbBr_3$. Therefore, our results show that the time scale, barrier to reorientation and the geometry of reorientational motions of the FA cation are affected significantly by changing the halide ion. This is in contrast with $MAPbX_3$ [34-35, 46], where the nature of the reorientational motion of the MA moiety did not show any significant variation with a change in the halide ion. In addition, MA cations undergo a 3-fold rotation about the C-N axis in the low temperature O phase and in the high temperature T and C phases an additional 4-fold reorientational jump of the whole molecule about an axis perpendicular to the C-N axis along with faster 3-fold rotation of MA cations around the C-N axis was observed. In $FAPbCl_3$, no signature of another jump mode could be observed across the O to T phase transition.

**Supporting Information:**

The descriptions of diffraction pattern and QENS spectra for $FAPbCl_3$, comparison between different possible models for reorientational motion of FA cation in orthorhombic, tetragonal and cubic phases and fits of HWHMs with $Q$ for $FAPbCl_3$ using fractional 2-fold jump rotation with respect to the C-H axis (in the orthorhombic phase) and fractional isotropic rotational diffusion (in the tetragonal and cubic phases) rotation model, are given in supporting information.

**Acknowledgements**

AM and DDS thank Science and Engineering Research Board, Nanomission, and Department of Science and Technology, Government of India, for support of this research. DDS thanks Jamsetji Tata Trust and CSIR Bhatnagar Fellowship for support. Access to the HFBS was provided by the Center for High Resolution Neutron Scattering, a partnership between the National Institute of Standards and Technology and the National Science Foundation under Agreement No. DMR-1508249.

**References**

1. Hao, F.; Stoumpos, C. C.; Cao, D. H.; Chang, R. P. H.; Kanatzidis, M. G., Lead-free solid-state organic–inorganic halide perovskite solar cells. *Nat. Photon.* **2014,** *8* (6), 489-494.

2. Kim, H.-S.; Lee, C.-R.; Im, J.-H.; Lee, K.-B.; Moehl, T.; Marchioro, A.; Moon, S.-J.; Humphry-Baker, R.; Yum, J.-H.; Moser, J. E.; Grätzel, M.; Park, N.-G., Lead Iodide Perovskite Sensitized All-Solid-State Submicron Thin Film Mesoscopic Solar Cell with Efficiency Exceeding 9%. *Sci. Rep.* **2012,** *2* (1), 591.

3. Dong, Q.; Fang, Y.; Shao, Y.; Mulligan, P.; Qiu, J.; Cao, L.; Huang, J., Electron-hole diffusion lengths > 175 μm in solution-grown CH3NH3PbI3 single crystals. *Science* **2015,** *347* (6225), 967-970.

4. Miyata, A.; Mitioglu, A.; Plochocka, P.; Portugall, O.; Wang, J. T.-W.; Stranks, S. D.; Snaith, H. J.; Nicholas, R. J., Direct measurement of the exciton binding energy and effective masses for charge carriers in organic–inorganic tri-halide perovskites. *Nat. Phys.* **2015,** *11* (7), 582-587.

5. *https://www.nrel.gov/pv/cell-efficiency.html*.

6. Xing, G.; Mathews, N.; Lim, S. S.; Yantara, N.; Liu, X.; Sabba, D.; Grätzel, M.; Mhaisalkar, S.; Sum, T. C., Low-temperature solution-processed wavelength-tunable perovskites for lasing. *Nat. Mater.* **2014,** *13* (5), 476-480.

7. Chen, Y.; Peng, J.; Su, D.; Chen, X.; Liang, Z., Efficient and Balanced Charge Transport Revealed in Planar Perovskite Solar Cells. *ACS App. Mater. Interfac.* **2015,** *7* (8), 4471-4475.

8. Kim, J. Y.; Lee, J.-W.; Jung, H. S.; Shin, H.; Park, N.-G., High-Efficiency Perovskite Solar Cells. *Chem. Rev.* **2020,** *120* (15), 7867-7918.

9. Stranks, S. D.; Snaith, H. J., Metal-halide perovskites for photovoltaic and light-emitting devices. *Nat. Nanotech.* **2015,** *10* (5), 391-402.

10. Saliba, M.; Matsui, T.; Seo, J.-Y.; Domanski, K.; Correa-Baena, J.-P.; Nazeeruddin, M. K.; Zakeeruddin, S. M.; Tress, W.; Abate, A.; Hagfeldt, A.; Grätzel, M., Cesium-containing triple cation perovskite solar cells: improved stability, reproducibility and high efficiency. *En. Environ. Sci.* **2016,** *9* (6), 1989-1997.

11. Liu, D.; Yang, J.; Kelly, T. L., Compact Layer Free Perovskite Solar Cells with 13.5% Efficiency. *J. Am. Chem. Soc.* **2014,** *136* (49), 17116-17122.

12. Wang, W.; Ma, Y.; Qi, L., High-Performance Photodetectors Based on Organometal Halide Perovskite Nanonets. *Adv. Funct. Mater.* **2017,** *27* (12), 1603653.

13. Rao, H.-S.; Li, W.-G.; Chen, B.-X.; Kuang, D.-B.; Su, C.-Y., In Situ Growth of 120 cm2 CH3NH3PbBr3 Perovskite Crystal Film on FTO Glass for Narrowband-Photodetectors. *Adv. Mater.* **2017,** *29* (16), 1602639.

14. Xiao, Z.; Kerner, R. A.; Zhao, L.; Tran, N. L.; Lee, K. M.; Koh, T.-W.; Scholes, G. D.; Rand, B. P., Efficient perovskite light-emitting diodes featuring nanometre-sized crystallites. *Nat. Photon.* **2017,** *11* (2), 108-115.

15. Shan, Q.; Li, J.; Song, J.; Zou, Y.; Xu, L.; Xue, J.; Dong, Y.; Huo, C.; Chen, J.; Han, B.; Zeng, H., All-inorganic quantum-dot light-emitting diodes based on perovskite emitters with low turn-on voltage and high humidity stability. *J. Mater. Chem. C* **2017,** *5* (18), 4565-4570.

16. Meng, L.; Yao, E.-P.; Hong, Z.; Chen, H.; Sun, P.; Yang, Z.; Li, G.; Yang, Y., Pure Formamidinium-Based Perovskite Light-Emitting Diodes with High Efficiency and Low Driving Voltage. *Adv. Mater.* **2017,** *29* (4), 1603826.

17. Veldhuis, S. A.; Boix, P. P.; Yantara, N.; Li, M.; Sum, T. C.; Mathews, N.; Mhaisalkar, S. G., Perovskite Materials for Light-Emitting Diodes and Lasers. *Adv. Mater.* **2016,** *28* (32), 6804-6834.

18. Kojima, A.; Teshima, K.; Shirai, Y.; Miyasaka, T., Organometal Halide Perovskites as Visible-Light Sensitizers for Photovoltaic Cells. *J. Am. Chem. Soc.* **2009,** *131* (17), 6050-6051.

19. Mozur, E. M.; Trowbridge, J. C.; Maughan, A. E.; Gorman, M. J.; Brown, C. M.; Prisk, T. R.; Neilson, J. R., Dynamical Phase Transitions and Cation Orientation-Dependent Photoconductivity in CH(NH2)2PbBr3. *ACS Mater. Lett.* **2019,** *1* (2), 260-264.

20. Shirzadi, E.; Mahata, A.; Roldán Carmona, C.; De Angelis, F.; Dyson, P. J.; Nazeeruddin, M. K., Introduction of a Bifunctional Cation Affords Perovskite Solar Cells Stable at Temperatures Exceeding 80 °C. *ACS Energy Lett.* **2019,** *4* (12), 2989-2994.

21. Miyata, K.; Atallah Timothy, L.; Zhu, X. Y., Lead halide perovskites: Crystal-liquid duality, phonon glass electron crystals, and large polaron formation. *Sci. Adv. 3* (10), e1701469.

22. Yang, Y.; Ostrowski, D. P.; France, R. M.; Zhu, K.; van de Lagemaat, J.; Luther, J. M.; Beard, M. C., Observation of a hot-phonon bottleneck in lead-iodide perovskites. *Nat. Photon.* **2016,** *10* (1), 53-59.

23. Pötz, W., Hot-phonon effects in bulk GaAs. *Phys. Rev. B* **1987,** *36* (9), 5016-5019.

24. Kim, M.; Im, J.; Freeman, A. J.; Ihm, J.; Jin, H., Switchable S = 1/2 and J = 1/2 Rashba bands in ferroelectric halide perovskites. *Proc. Nat. Acad. Sci.* **2014,** *111* (19), 6900.

25. Nagai, M.; Tomioka, T.; Ashida, M.; Hoyano, M.; Akashi, R.; Yamada, Y.; Aharen, T.; Kanemitsu, Y., Longitudinal Optical Phonons Modified by Organic Molecular Cation Motions in Organic-Inorganic Hybrid Perovskites. *Phys. Rev. Lett.* **2018,** *121* (14), 145506.

26. Li, J.; Rinke, P., Atomic structure of metal-halide perovskites from first principles: The chicken-and-egg paradox of the organic-inorganic interaction. *Phys. Rev. B* **2016,** *94* (4), 045201.

27. Yaffe, O.; Guo, Y.; Tan, L. Z.; Egger, D. A.; Hull, T.; Stoumpos, C. C.; Zheng, F.; Heinz, T. F.; Kronik, L.; Kanatzidis, M. G.; Owen, J. S.; Rappe, A. M.; Pimenta, M. A.; Brus, L. E., Local Polar Fluctuations in Lead Halide Perovskite Crystals. *Phys. Rev. Lett.* **2017,** *118* (13), 136001.

28. Sharma, R.; Dai, Z.; Gao, L.; Brenner, T. M.; Yadgarov, L.; Zhang, J.; Rakita, Y.; Korobko, R.; Rappe, A. M.; Yaffe, O., Elucidating the atomistic origin of anharmonicity in tetragonal CH3NH3PbI3 with Raman scattering. *Phys. Rev. Mater.* **2020,** *4* (9), 092401.

29. Leguy, A. M. A.; Goñi, A. R.; Frost, J. M.; Skelton, J.; Brivio, F.; Rodríguez-Martínez, X.; Weber, O. J.; Pallipurath, A.; Alonso, M. I.; Campoy-Quiles, M.; Weller, M. T.; Nelson, J.; Walsh, A.; Barnes, P. R. F., Dynamic disorder, phonon lifetimes, and the assignment of modes to the vibrational spectra of methylammonium lead halide perovskites. *Phys. Chem. Chem. Phys.* **2016,** *18* (39), 27051-27066.

30. Johnston, A.; Walters, G.; Saidaminov, M. I.; Huang, Z.; Bertens, K.; Jalarvo, N.; Sargent, E. H., Bromine Incorporation and Suppressed Cation Rotation in Mixed-Halide Perovskites. *ACS Nano* **2020,** *14* (11), 15107-15118.

31. Kubicki, D. J.; Prochowicz, D.; Hofstetter, A.; Péchy, P.; Zakeeruddin, S. M.; Grätzel, M.; Emsley, L., Cation Dynamics in Mixed-Cation (MA)x(FA)1–xPbI3 Hybrid Perovskites from Solid-State NMR. *J. Am. Chem. Soc.* **2017,** *139* (29), 10055-10061.

32. Baikie, T.; Barrow, N. S.; Fang, Y.; Keenan, P. J.; Slater, P. R.; Piltz, R. O.; Gutmann, M.; Mhaisalkar, S. G.; White, T. J., A combined single crystal neutron/X-ray diffraction and solid-state nuclear magnetic resonance study of the hybrid perovskites CH3NH3PbX3 (X = I, Br and Cl). *J. Mater. Chem. A* **2015,** *3* (17), 9298-9307.

33. Bakulin, A. A.; Selig, O.; Bakker, H. J.; Rezus, Y. L. A.; Müller, C.; Glaser, T.; Lovrincic, R.; Sun, Z.; Chen, Z.; Walsh, A.; Frost, J. M.; Jansen, T. L. C., Real-Time Observation of Organic Cation Reorientation in Methylammonium Lead Iodide Perovskites. *J. Phys. Chem. Lett.* **2015,** *6* (18), 3663-3669.

34. Sharma, V. K.; Mukhopadhyay, R.; Mohanty, A.; Sakai, V. G.; Tyagi, M.; Sarma, D. D., Contrasting Effects of FA Substitution on MA/FA Rotational Dynamics in FAxMA1–xPbI3. *J. Phys. Chem. C* **2021,** *125* (24), 13666-13676.

35. Sharma, V. K.; Mukhopadhyay, R.; Mohanty, A.; Tyagi, M.; Embs, J. P.; Sarma, D. D., Contrasting Behaviors of FA and MA Cations in APbBr3. *J. Phys. Chem. Lett.* **2020,** *11* (22), 9669-9679.

36. Schuck, G.; Lehmann, F.; Ollivier, J.; Mutka, H.; Schorr, S., Influence of Chloride Substitution on the Rotational Dynamics of Methylammonium in MAPbI3–xClx Perovskites. *J. Phys. Chem. C* **2019,** *123* (18), 11436-11446.

37. Li, B.; Kawakita, Y.; Liu, Y.; Wang, M.; Matsuura, M.; Shibata, K.; Ohira-Kawamura, S.; Yamada, T.; Lin, S.; Nakajima, K.; Liu, S., Polar rotor scattering as atomic-level origin of low mobility and thermal conductivity of perovskite CH3NH3PbI3. *Nat. Commun.* **2017,** *8* (1), 16086.

38. Koegel, A. A.; Mozur, E. M.; Oswald, I. W. H.; Jalarvo, N. H.; Prisk, T. R.; Tyagi, M.; Neilson, J. R., Correlating Broadband Photoluminescence with Structural Dynamics in Layered Hybrid Halide Perovskites. *Journal of the American Chemical Society* **2022,** *144* (3), 1313-1322.

39. Grüninger, H.; Bokdam, M.; Leupold, N.; Tinnemans, P.; Moos, R.; De Wijs, G. A.; Panzer, F.; Kentgens, A. P. M., Microscopic (Dis)order and Dynamics of Cations in Mixed FA/MA Lead Halide Perovskites. *J. Phys. Chem. C* **2021,** *125* (3), 1742-1753.

40. Mattoni, A.; Filippetti, A.; Caddeo, C., Modeling hybrid perovskites by molecular dynamics. *J. Phys.: Cond. Matt.* **2016,** *29* (4), 043001.

41. Lahnsteiner, J.; Kresse, G.; Kumar, A.; Sarma, D. D.; Franchini, C.; Bokdam, M., Room-temperature dynamic correlation between methylammonium molecules in lead-iodine based perovskites: An ab initio molecular dynamics perspective. *Phys. Rev. B* **2016,** *94* (21), 214114.

42. Govinda, S.; Kore, B. P.; Bokdam, M.; Mahale, P.; Kumar, A.; Pal, S.; Bhattacharyya, B.; Lahnsteiner, J.; Kresse, G.; Franchini, C.; Pandey, A.; Sarma, D. D., Behavior of Methylammonium Dipoles in MAPbX3 (X = Br and I). *J. Phys. Chem. Lett.* **2017,** *8* (17), 4113-4121.

43. Bokdam, M.; Lahnsteiner, J.; Sarma, D. D., Exploring Librational Pathways with on-the-Fly Machine-Learning Force Fields: Methylammonium Molecules in MAPbX3 (X = I, Br, Cl) Perovskites. *J. Phys. Chem. C* **2021,** *125* (38), 21077-21086.

44. Li, J.; Järvi, J.; Rinke, P., Multiscale model for disordered hybrid perovskites: The concept of organic cation pair modes. *Phys. Rev. B* **2018,** *98* (4), 045201.

45. Brown, K. L.; Parker, S. F.; García, I. R.; Mukhopadhyay, S.; Sakai, V. G.; Stock, C., Molecular orientational melting within a lead-halide octahedron framework: The order-disorder transition in CH33NH3PbBr3. *Physical Review B* **2017,** *96* (17), 174111.

46. Songvilay, M.; Wang, Z.; Sakai, V. G.; Guidi, T.; Bari, M.; Ye, Z. G.; Xu, G.; Brown, K. L.; Gehring, P. M.; Stock, C., Decoupled molecular and inorganic framework dynamics in CH3NH3PbCl3. *Physical Review Materials* **2019,** *3* (12), 125406.

47. Leguy, A. M. A.; Frost, J. M.; McMahon, A. P.; Sakai, V. G.; Kockelmann, W.; Law, C.; Li, X.; Foglia, F.; Walsh, A.; O'Regan, B. C.; Nelson, J.; Cabral, J. T.; Barnes, P. R. F., The dynamics of methylammonium ions in hybrid organic–inorganic perovskite solar cells. *Nature Communications* **2015,** *6* (1), 7124.

48. Swainson, I. P.; Stock, C.; Parker, S. F.; Van Eijck, L.; Russina, M.; Taylor, J. W., From soft harmonic phonons to fast relaxational dynamics in ${\mathrm{CH}}_{3}{\mathrm{NH}}_{3}{\mathrm{PbBr}}_{3}$. *Phys. Rev. B* **2015,** *92* (10), 100303.

49. Eperon, G. E.; Stranks, S. D.; Menelaou, C.; Johnston, M. B.; Herz, L. M.; Snaith, H. J., Formamidinium lead trihalide: a broadly tunable perovskite for efficient planar heterojunction solar cells. *Energy Environ. Sci.* **2014,** *7* (3), 982-988.

50. Han, Q.; Bae, S.-H.; Sun, P.; Hsieh, Y.-T.; Yang, Y.; Rim, Y. S.; Zhao, H.; Chen, Q.; Shi, W.; Li, G., Single Crystal Formamidinium Lead Iodide (FAPbI3): Insight into the Structural, Optical, and Electrical Properties. *Adv. Mater.* **2016,** *28* (11), 2253-2258.

51. Lee, J.-W.; Seol, D.-J.; Cho, A.-N.; Park, N.-G., High-Efficiency Perovskite Solar Cells Based on the Black Polymorph of HC(NH2)2PbI3. *Adv. Mater.* **2014,** *26* (29), 4991-4998.

52. Bohn, B. J.; Tong, Y.; Gramlich, M.; Lai, M. L.; Döblinger, M.; Wang, K.; Hoye, R. L. Z.; Müller-Buschbaum, P.; Stranks, S. D.; Urban, A. S.; Polavarapu, L.; Feldmann, J., Boosting Tunable Blue Luminescence of Halide Perovskite Nanoplatelets through Postsynthetic Surface Trap Repair. *Nano Letters* **2018,** *18* (8), 5231-5238.

53. Kubicki, D. J.; Saski, M.; MacPherson, S.; Gałkowski, K.; Lewiński, J.; Prochowicz, D.; Titman, J. J.; Stranks, S. D., Halide Mixing and Phase Segregation in Cs2AgBiX6 (X = Cl, Br, and I) Double Perovskites from Cesium-133 Solid-State NMR and Optical Spectroscopy. *Chemistry of Materials* **2020,** *32* (19), 8129-8138.

54. Zhang, H.; Lv, Y.; Wang, J.; Ma, H.; Sun, Z.; Huang, W., Influence of Cl Incorporation in Perovskite Precursor on the Crystal Growth and Storage Stability of Perovskite Solar Cells. *ACS Applied Materials & Interfaces* **2019,** *11* (6), 6022-6030.

55. Wang, J.; Peng, J.; Sun, Y.; Liu, X.; Chen, Y.; Liang, Z., FAPbCl3 Perovskite as Alternative Interfacial Layer for Highly Efficient and Stable Polymer Solar Cells. *Adv. Electron. Mater.* **2016,** *2* (11), 1600329.

56. Lv, S.; Pang, S.; Zhou, Y.; Padture, N. P.; Hu, H.; Wang, L.; Zhou, X.; Zhu, H.; Zhang, L.; Huang, C.; Cui, G., One-step, solution-processed formamidinium lead trihalide (FAPbI(3−x)Clx) for mesoscopic perovskite–polymer solar cells. *Phys. Chem. Chem. Phys.* **2014,** *16* (36), 19206-19211.

57. Parfenov, A. A.; Yamilova, O. R.; Gutsev, L. G.; Sagdullina, D. K.; Novikov, A. V.; Ramachandran, B. R.; Stevenson, K. J.; Aldoshin, S. M.; Troshin, P. A., Highly sensitive and selective ammonia gas sensor based on FAPbCl3 lead halide perovskites. *J. Mater. Chem. C* **2021,** *9* (7), 2561-2568.

58. Mączka, M.; Ptak, M.; Vasconcelos, D. L. M.; Giriunas, L.; Freire, P. T. C.; Bertmer, M.; Banys, J.; Simenas, M., NMR and Raman Scattering Studies of Temperature- and Pressure-Driven Phase Transitions in CH3NH2NH2PbCl3 Perovskite. *The Journal of Physical Chemistry C* **2020,** *124* (49), 26999-27008.

59. Bee, M., Quasielastic neutron scattering. *Adam Hilger: United Kingdom* **1988**.

60. Meyer, A.; Dimeo, R. M.; Gehring, P. M.; Neumann, D. A.; H., M.-L., The high-flux backscattering spectrometer at the NIST Center for Neutron Research. *Review of Scientific Instruments* **2003,** *74* (5), 2759-2777.

61. Azuah, R. T. K., L. R.; Qiu, Y.; Tregenna-Piggott, P. L.; Brown, C. M.; Copley, J. R.; Dimeo, R. M., DAVE: A Comprehensive Software Suite for the Reduction, Visualization, and Analysis of Low Energy Neutron Spectroscopic Data. *J. Res. Natl. Inst. Stand. Technol.* **2009,** *114*, 341.

62. Carlile, C. J.; Adams, M. A., The design of the IRIS inelastic neutron spectrometer and improvements to its analysers. *Physica B: Condensed Matter* **1992,** *182* (4), 431-440.

63. Taylor, J.; Arnold, O.; Bilheaux, J.; Buts, A.; Campbell, S.; Doucet, M.; Draper, N.; Fowler, R.; Gigg, M.; Lynch, V.; Markvardsen, A.; Palmen, K.; Parker, P.; Peterson, P.; Ren, S.; Reuter, M.; Savici, A.; Taylor, R.; Tolchenov, R.; Whitley, R.; Zhou, W.; Zikovsky, J., Mantid, A high performance framework for reduction and analysis of neutron scattering data. 2012; Vol. 2012, p W26.010.

64. Gallop, N. P.; Selig, O.; Giubertoni, G.; Bakker, H. J.; Rezus, Y. L. A.; Frost, J. M.; Jansen, T. L. C.; Lovrincic, R.; Bakulin, A. A., Rotational Cation Dynamics in Metal Halide Perovskites: Effect on Phonons and Material Properties. *The Journal of Physical Chemistry Letters* **2018,** *9* (20), 5987-5997.

65. Sears, V. F., THEORY OF COLD NEUTRON SCATTERING BY HOMONUCLEAR DIATOMIC LIQUIDS: I. FREE ROTATION. *Canadian Journal of Physics* **1966,** *44* (6), 1279-1297.

66. Govinda, S.; Kore, B. P.; Swain, D.; Hossain, A.; De, C.; Guru Row, T. N.; Sarma, D. D., Critical Comparison of FAPbX3 and MAPbX3 (X = Br and Cl): How Do They Differ? *The Journal of Physical Chemistry C* **2018,** *122* (25), 13758-13766.

67. Schueller, E. C.; Laurita, G.; Fabini, D. H.; Stoumpos, C. C.; Kanatzidis, M. G.; Seshadri, R., Crystal Structure Evolution and Notable Thermal Expansion in Hybrid Perovskites Formamidinium Tin Iodide and Formamidinium Lead Bromide. *Inorganic Chemistry* **2018,** *57* (2), 695-701.

68. Keshavarz, M.; Ottesen, M.; Wiedmann, S.; Wharmby, M.; Küchler, R.; Yuan, H.; Debroye, E.; Steele, J. A.; Martens, J.; Hussey, N. E.; Bremholm, M.; Roeffaers, M. B. J.; Hofkens, J., Tracking Structural Phase Transitions in Lead-Halide Perovskites by Means of Thermal Expansion. *Advanced Materials* **2019,** *31* (24), 1900521.

69. A. Franz, D. M. T., F. Lehmann, M. Kärgell and S. Schorr, The influence of deuteration on the crystal structure of hybrid halide perovskites: a temperature-dependent neutron diffraction study of FAPbBr$ \sb 3 $. *Acta Crystallographica Section B* **2020,** *76* (2), 267--274.

# Supporting Information

It is reported by Govinda et al[1] that FAPbCl$_3$ undergoes two first order phase transitions at ca 258 K and 271 K as observed in DSC measurements. It is also shown that at low temperatures (200 K and 100 K) the system exists in orthorhombic (O) structure and at room temperature (295K) it is in cubic (C) phase. However, the structure in the region between these two transitions i.e. 258K to 271K was not reported and no such literature is available for this phase. As mentioned in the manuscript, by virtue of the presence of a small number of diffraction detectors in the IRIS spectrometer, diffraction patterns could be recorded simultaneously to the QENS data. However, these diffraction patterns are of limited $d$-spacing range. The neutron diffraction patterns for FAPbCl$_3$ so obtained are shown in Fig. S1 at three different temperatures, 250K (below the first transition), 265K (in between the two phase transitions) and 280 K (above the second transition). The data clearly show the presence of three distinct phases at these temperatures. We have used these diffraction patterns to specify the structure of the intermediate phase (258 – 271 K). First, we have used the space group and lattice parameters obtained from the ref. 1 for cubic and orthorhombic phase and refined the lattice parameter using FullProf. Albeit not a full refinement due to the limited range accessed by IRIS, the observed diffraction patterns could be described well (Fig. S1), and the obtained lattice parameters (Table-S1) are found to be consistent with the earlier reported results[1]. The $APbX_3$ ($A$=MA, FA) family of compounds, in general, forms orthorhombic structure at low temperature which goes to cubic phase via tetragonal phase[1-3]. As there is no information available for the intermediate phase, we attempted to obtain the same using the structural parameters as reported for the tetragonal phase of FAPbBr$_3$ [2] and refined the data, for the lattice parameters only. We find that at 265 K it is most likely to be a tetragonal structure (Fig. S1).

Table S1: Structural parameters obtained from the refinement of diffraction patterns

| T (K) | Crystal system | Space group | Lattice (Å) | Volume(Å$^3$) |
|---|---|---|---|---|
| 250K | Orthorohmbic | Cmcm | a=8.8266<br>b=7.4384<br>c= 11.4783 | 753.6161 |
| 265K | Tetragonal | P4/mbm | a=7.9077<br>b=7.9077<br>c= 6.5112 | 407.1563 |
| 280K | Cubic | P m-3m | a=5.7536 | 190.4706 |

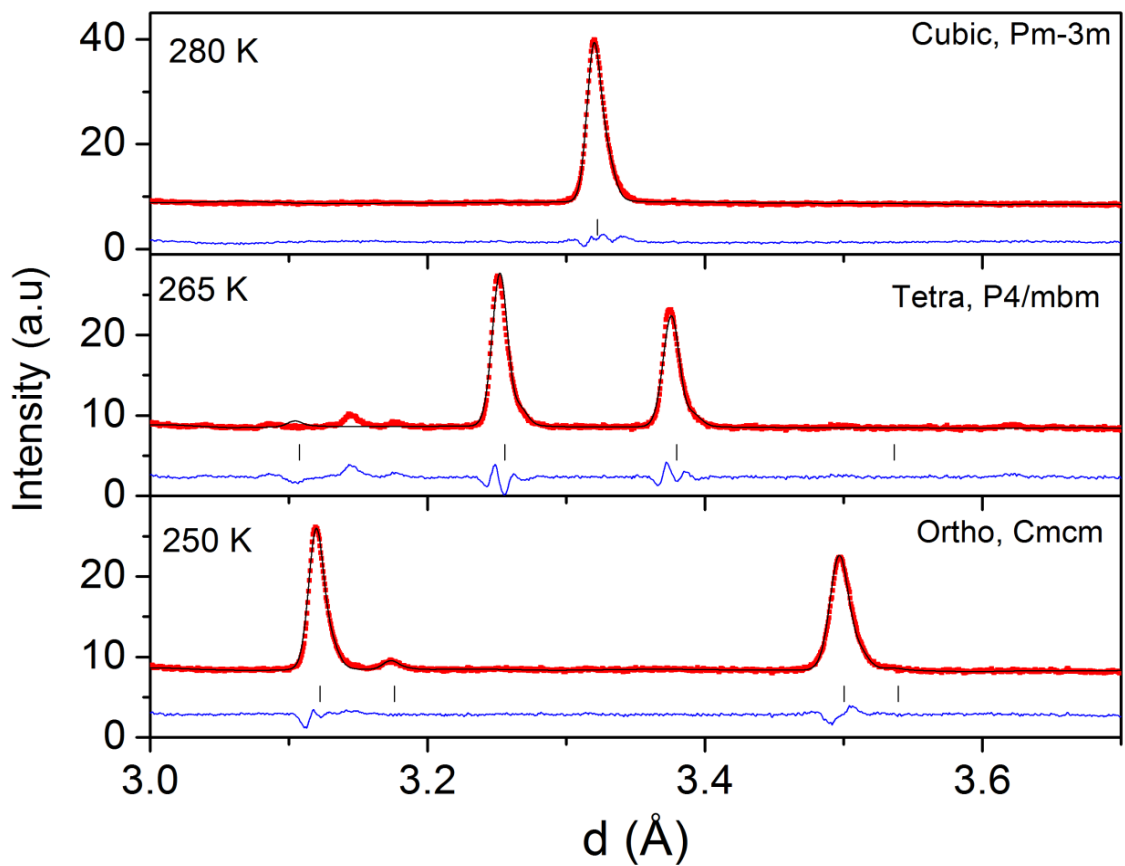

Fig. S1 The diffraction patterns for FAPbCl$_3$ recorded using IRIS spectrometer at 250K, 265K and 280K. Solid lines are the fits assuming orthorhombic (Cmcm), tetragonal (P4/mbm) and cubic (Pm-3m) phases [1,2].

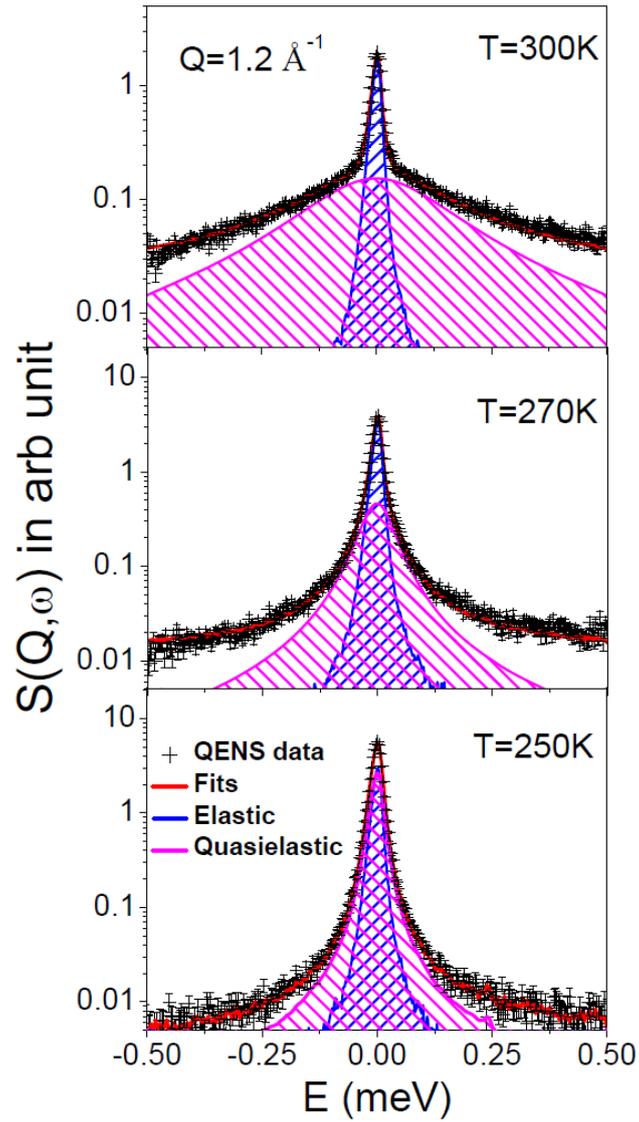

Fig. S2: Typical fitted QENS spectra ($Q = 1.2$ Å$^{-1}$) for FAPbCl$_3$ at 250, 270 and 300 K where system is in orthorhombic, tetragonal and cubic phases, respectively. Elastic (blue) and quasielastic (magenta) components are also shown.

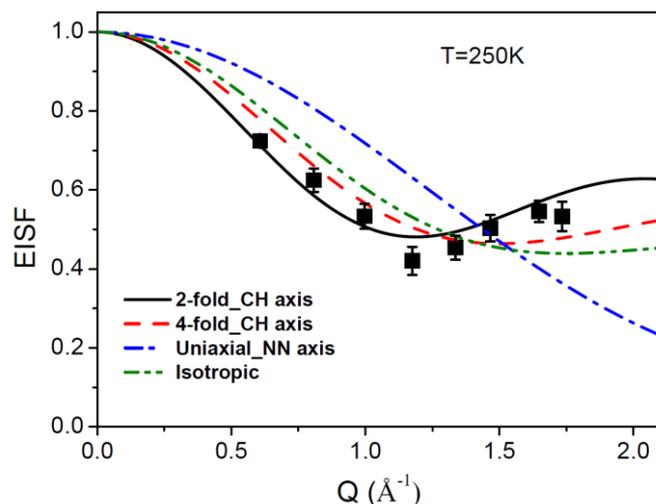

Fig. S3 Comparison of various possible models for the rotational motion of FA cation in orthorhombic phase at a typical temperature of 250 K. Solid (black), dashed (red), dash-dotted (blue) and dash double dotted (green) lines correspond to the least-squares fits assuming, 2-fold jump rotation about the C−H axis, 4-fold jump rotation about the C−H axis, uniaxial rotation around the N−N axis, and isotropic rotational diffusion models. It is evident that the 2-fold jump rotation about the C−H axis with about 88% mobile FA cations describes the experimental EISF the best.

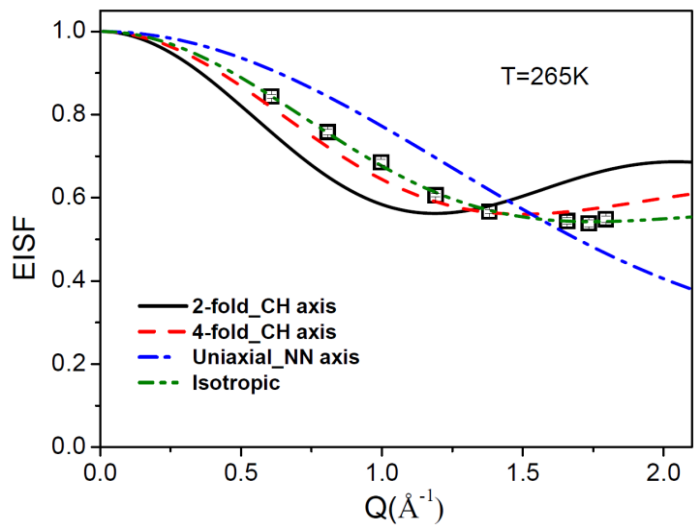

Fig. S4 Comparison of various possible models for the rotational motion of FA cation in tetragonal phase at a typical temperature of 265 K. Solid (black), dashed (red), dash-dotted (blue) and dash double dotted (green) lines correspond to the least-squares fits assuming, 2-fold jump rotation about the C−H axis, 4-fold jump rotation about the C−H axis, uniaxial rotation around the N−N axis, and isotropic rotational diffusion models. It is evident that isotropic rotational diffusion with 46% mobile FA cations describes the experimental EISF best.

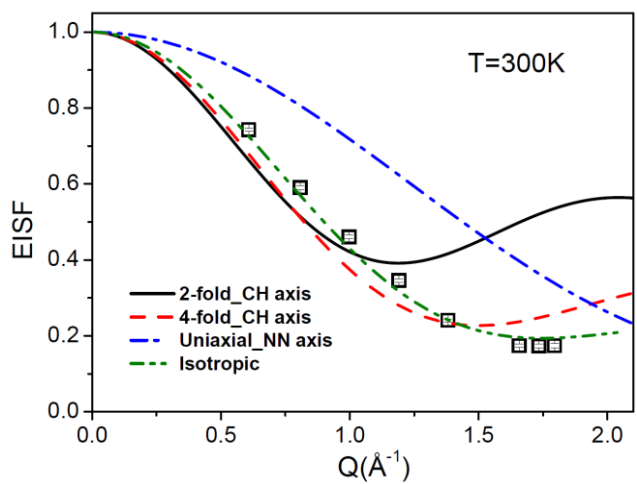

Fig. S5 Comparison of various possible models for the FA cation reorientation in the cubic phase at a typical temperature of 300 K. Solid (black), dashed (red), dash-dotted (blue) and dash double dotted (green) lines correspond to the least-squares fits assuming, 2-fold jump rotation about the C−H axis, 4-fold jump rotation about the C−H axis, uniaxial rotation around the N−N axis, and isotropic rotational diffusion models. It is evident that isotropic rotational diffusion with 81% mobile FA cations describes the experimental EISF best.

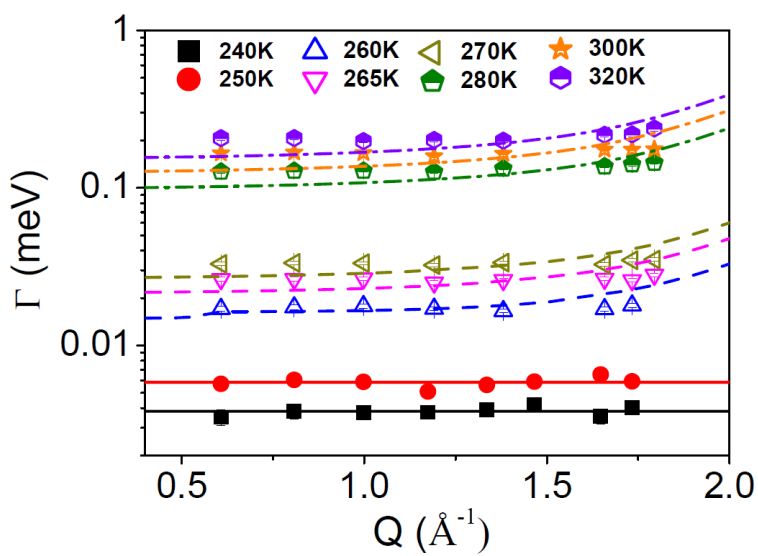

Fig.S6 Variation of HWHM for $FAPbCl_3$ with $Q$ at different temperatures. It is evident that HWHM shows step jumps around both the crystallographic orthorhombic-tetragonal and tetragonal-cubic phase transitions. Solid and dashed lines are as per the fits assuming fractional 2-fold jump rotation with respect to the C-H axis (in the orthorhombic phase) and fractional isotropic rotational diffusion (in the tetragonal and cubic phases) models, respectively.

**References**

1. Govinda, S.; Kore, B. P.; Swain, D.; Hossain, A.; De, C.; Guru Row, T. N.; Sarma, D. D., Critical Comparison of FAPbX3 and MAPbX3 (X = Br and Cl): How Do They Differ? *The Journal of Physical Chemistry C* **2018,** *122* (25), 13758-13766.

2. A. Franz, D. M. T., F. Lehmann, M. Kärgell and S. Schorr, The influence of deuteration on the crystal structure of hybrid halide perovskites: a temperature-dependent neutron diffraction study of FAPbBr$_3$. *Acta Crystallographica Section B* **2020,** *76* (2), 267--274.

3. Poglitsch, A.; Weber, D. Dynamic disorder in methylammoniumtrihalogenoplumbates (II) observed by millimeter-wave spectroscopy. *J. Chem. Phys.* **1987**, 87 (11), 6373−6378.